# Improve photon number discrimination for a superconducting series nanowire detector by applying a digital matched filter

Hao Hao[1], Qing-Yuan Zhao[1,2]*, Ling-Dong Kong[1], Shi Chen[1], Hui Wang[1], Yang-Hui Huang[1], Jia-Wei Guo[1], Wan Chao[2], Hao Liu[2], Xue-Cou Tu[1,2], La-Bao Zhang[1,2], Xiao-Qing Jia[1,2], Jian Chen[1,2], Lin Kang[1,2], Cong Li[3], Te Chen[3], Gui-Xing Cao[3], and Pei-Heng Wu[1,2]

[1]*Research Institute of Superconductor Electronics (RISE), School of Electronic Science and Engineering, Nanjing University, Nanjing, Jiangsu 210023, China*

[2]*Purple Mountain Laboratories, Nanjing, Jiangsu 211111, China*

[3]*Institute of Telecommunication and Navigation Satellites, China Academy of Space Technology, Beijing 100094, China*

*qyzhao@nju.edu.cn.

## Abstract:

Photon number resolving (PNR) is an important capacity for detectors working in quantum and classical applications. Although a conventional superconducting nanowire single-photon detector (SNSPD) is not a PNR detector, by arranging nanowires in a series array and multiplexing photons over space, such series PNR-SNSPD can gain quasi-PNR capacity. However, the accuracy and maximum resolved photon number are both limited by the signal-to-noise (SNR) ratio of the output pulses. Here, we introduce a matched filter, which is an optimal filter in terms of SNR for SNSPD pulses. Experimentally, compared to conventional readout using a room-temperature amplifier, the normalized spacing between pulse amplitudes from adjacent photon number detections increased by a maximum factor of 2.1 after the matched filter. Combining with a cryogenic amplifier to increase SNR further, such spacing increased by a maximum factor of 5.3. In contrast to a low pass filter, the matched filter gave better SNRs while maintaining good timing jitters. Minimum timing jitter of 55 ps was obtained experimentally. Our results suggest that the matched filter is a useful tool for improving the performance of the series PNR-SNSPD and the maximum resolved photon number can be expected to reach 65 or even large.

## Novelty:

Photon number resolving (PNR) is an important capacity for detectors working in quantum and classical applications. However, the dynamic range of a PNR detector is limited by the signal-to-noise ratio of electrical readout. Here, we designed an optimal matched filter on a series superconducting nanowire single-photon detector (SNSPD). Experimental results showed that the SNR of a 6-pixel PNR-SNSPD was improved by 3~4 times while the timing jitter was maintained to be 55 ps. Our results suggest that the matched filter is a useful tool for improving the performance of the series PNR-SNSPD and the maximum resolved photon number can be expected to reach 65 or even large.

# Main article:

Superconducting nanowire single-photon detectors (SNSPDs) exhibits great performance, such as near unitary detection efficiency, ultralow dark counts, sub-5ps timing jitter, and high counting rate[1] [2]. In an SNSPD, a photon absorption triggers a voltage pulse of fixed amplitude, which is independent of the photon number or photon energy information. Therefore, a conventional SNSPD designed into a single meandered nanowire is not capable of resolving the photon number. As the photon number state (Fock state) is an important quantum state and it is the basis for a coherent state, a photon number resolving (PNR) detector is needed in many quantum photonic experiments. For example, a PNR detector is one of the key elements in the linear optical quantum computing (LOQC) architecture[3]. With a PNR detector for reconstructing the coherent state, a quantum key distribution (QKD) can gain a higher key generation rate[4] and a quantum communication receiver can be built to overcome the standard quantum limit[5]. For classical applications, such as free space communication[6], LiDAR[7], and imaging[8], a PNR detector increases the dynamic range of response, which certainly can improve the signal-to-noise ratio (SNR) of the receiver when the incident light is beyond the single-photon regime.

As the superconducting transition is localized within a short distance around the initial hotspot generated after a photon absorption, a quasi-PNR capacity can be achieved by spatially multiplexing photons and detecting with a large nanowire array. There are two types PNR-SNSPDs. One is connecting nanowires in parallel[9] [10] while the other uses a series structure[11]. The series PNR-SNSPD can be biased close to the critical current. Thus, it typically can have higher detection efficiency[12] and a large dynamic range[13]. For a series PNR-SNSPD, the output signal is about $V_\text{n} \approx \frac{n \cdot I_\text{B} \cdot Z_0}{N + Z_0/R_\text{s}}$, where $n$ is the number of fired nanowires, $I_\text{B}$ is the bias current, $N$ is the array size, $Z_0$ is the characteristic impedance of the readout amplifier and $R_\text{s}$ is the shunted resistance. Typically, $Z_0$ for a commercial RF amplifier is designed to be 50 Ω to match the impedance of a coaxial cable and $R_\text{S}$ is around $10\Omega \sim 50\ \Omega$ (in our experiment $R_s$

≅ 52 Ω). Thus, as the array size $N$ increases, $V_n$ dropped proportional to $1/N$, resulting in a worse SNR for distinguishing the pulse amplitude. Consequently, it limits the resolution for accurately discriminating different photon number states and the maximum photon number that can be resolved. So far, the maximum resolved photon number reported for a series PNR-SNSPD was 24[13].

To overcome the above tradeoff between accurate photon number discrimination and the maximum photon number, the SNR of the series PNR-SNSPD needs to be improved. A high impedance RF amplifier with $Z_0 \gg R_S$ seems an ideal solution. However, at the cryogenic temperature, it is difficult to build a high speed, high impedance, and low noise amplifier[14]. To avoid parasitic capacitance, a high-impedance FET transistor has to be put close to the detector. However, as the power dissipation of a typical transistor is above mW, such close integration would elevate the local temperature of the detector. Placing a cryogenic amplifier on a different temperature stage from detectors is more feasible. However, as the coaxial cables typically have 50 Ω impedance, the impedance of a cryogenic amplifier needs to be 50 Ω as well to avoid reflections. In addition, designing the nanowire detector into avalanche structure[15] or choosing superconductors of higher critical current density can output a higher signal pulse. However, these modifications of the nanowire need to be done carefully so that the detection performance, such as timing jitter and detection efficiency, would not be deteriorated.

Here, instead of upgrading the readout setup or modifying the detector design, we introduce a post signal processing approach for improving the discrimination of a series PNR-SNSPD. Since the processing is done after the digitization of the detection pules, this method can be implemented in the existing system straightforwardly. We designed a digital match filter and used the location and amplitude of the peak, where the SNR is the maximum, to extract the information of photon number and photon arrival time, respectively. Compared to a RT amplifier readout, the normalized spacing between adjacent photon detection pulse amplitude increased by a factor of 2.1 for maximum

(1.8 on average) after the matched filter. Combined with a cryogenic amplifier, such normalized spacing further increased by a factor of 5.3 for maximum (4.1 on average). Meanwhile, the time jitter was reduced from 94 ps to 86 ps for single-photon detections and from 73 ps to 55 ps for five-photon detections. We also compared the performance of the matched filter with a low-pass filter. Although a low-pass filter can improve the SNR by setting a low cut-off frequency, due to the loss of the high-frequency components, the timing jitter got worse. In contrast, a matched filter can simultaneously improve SNR and maintain low timing jitter, letting it be a promising signal processing method for SNSPD and other nanowire-based devices, e.g. nanowire cryotrons[16].

We first analyze the signal property of the output pulse from a series PNR-SNSPD in both time domain and frequency domain, based which we can select a proper filtering strategy. The time-domain response $r_n(t)$ can be written as the superposition of the signal $s_n(t)$ and the noise $g(t)$, where $n$ is the number of fired nanowires.

$$r_n(t) = s_n(t) + g(t) \tag{1}$$

$s_n(t)$ is the response function of the $n$-photon detection, which can be fitted by a two-component bi-exponential model as shown in Equ.2[17].

$$s_n(t) = A_n \cdot (e^{-t/\tau_1} - e^{-t/\tau_2}) \tag{2}$$

$\tau_1$ and $\tau_2$ represent the time constants for the rising edge and the falling edge. By fitting our experimental data for our 6-pixel PNR-SNSPD[12], we have $\tau_1 = 0.21$ ns and $\tau_2 = 25$ ns. $A_n$ is the amplitude of $n$-photon detection pulses. We set $A_n$ as the average value from measurement data to include the nonlinear dependence of $A_n$ on $n$. We assumed $g(t)$ as a stationary Gaussian white noise with a standard deviation of $\sigma = 0.40$ mV. The SNR for the $n$-photon detection pulse is then

$$SNR_n = A_n/\sigma \tag{3}$$

Fig.1a simulates the output waveforms of our 6-pixels SNSPD by using Equ.1 and 2 with taking experimental parameters. It clearly shows that in the presence of noise accurate discrimination of different $n$-photon pulses becomes difficult. We also did a Monte Carlo simulation to extract the $SNR_n$. The data is labeled in Fig.1a as well.

By applying a Fourier transfer of the time domain signals, we can have the frequency spectrum of the *n*-photon detection pulses. As shown in Fig.1b, most of the power in the signal spectrum *S(f)* stays at low frequencies, while the noise has a flat spectrum *G(f)*. We define the characteristic frequency $f_c$ as the intersection of $S(f_c) = G(f_c)$. With a cryogenic amplifier, we have $f_c$ = 65 MHz. Only for frequency components lower than $f_c$, the SNR in the frequency domain is larger than 1. It indicates that a low-pass filter of cutoff frequency lower than $f_c$ can improve SNR. However, as the high-frequency components construct the sharp rising edge, if the cutoff frequency is set too low, it will scarify the sharpness of the rising edge and deteriorate the timing accuracy[18].

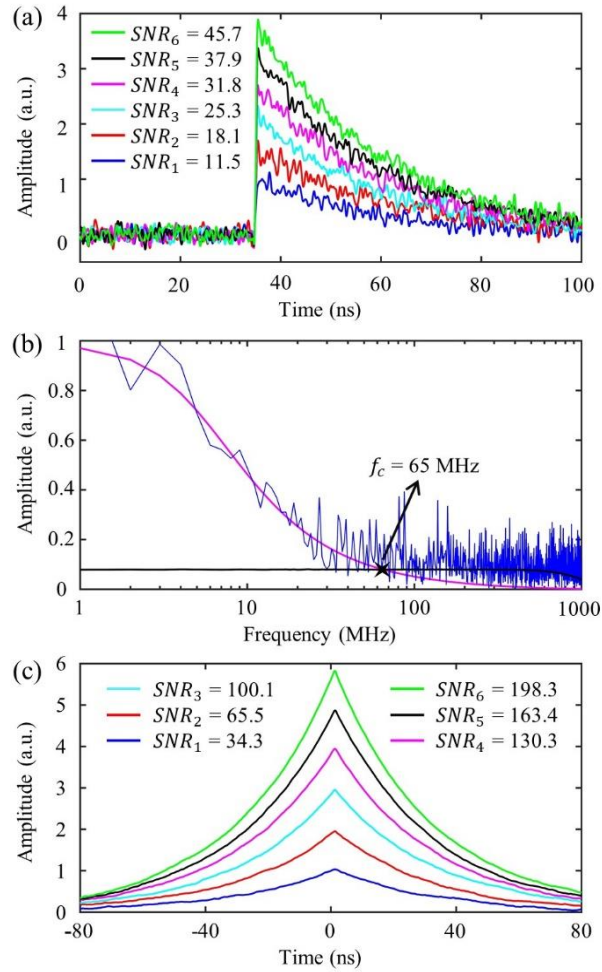

FIG. 1. (a) Simulated photon detection pulses $r_n(t)$ with experimental parameters using Equ.1 and 2. (b) The frequency spectrum for a pure signal *s(t)* (pink), a Gaussian white noise *g(t)* (black) blue, and an experimental pulse (blue) read by a cryogenic amplifier. The star marks the characteristic frequency $f_c$. (c) The output pulse $y_n(t)$ after a matched filter. The $SNR_n$ values shown in (a) and (c), which are defined as the ratio between the peak value of $y_n(t)$

and its standard deviation, are extracted from a Monte Carlo simulation. Amplitude values are normalized to the single-photon detections.

From the above analysis, we can conclude two basic properties of the output for a series PNR-SNSPD. First, the time domain waveform of the output pulse has a constant shape which can be well characterized by using Equ.1 and 2. Second, the frequency spectrum of the output pulse is wideband and non-uniform, indicating that simply filtering within a certain frequency range is not optimal. These properties inspire us to implement matched filters. In signal processing, matched filters are a basic tool in electrical engineering for extracting known wavelets from a signal that has been contaminated by noise. This is accomplished by cross-correlating the signal with the wavelet, which is equivalent to convolving the unknown signal with a conjugated time-reversed version of the template. The matched filter is the optimal linear filter for maximizing the signal-to-noise ratio (SNR) in the presence of additive stochastic noise[19].

In our case, the impulse response of the matched filter is $h(t) = s_n(t_0 - t)$, where $t_0$ is the duration of the filter. Then, the filtering process is done by

$$y_n(t) = r_n(t) \otimes s_n(t_0 - t) \qquad (4)$$

where $\otimes$ is the convolution operator. As shown in Fig. 1c, by applying a matched filter, the filtered output $y_n$ shows a much better SNR. Since the best SNR after a matched filter occurs when the signal overlaps with the wavelet, namely at the peak value $p_n$ of $y_n$, we use $p_n$ to distinguish photon number and the time of the peak $t_{pn}$ to define the photon arrival times. By running a Monte Carlo simulation, we extract the SNR values of $y_n$. As labeled in Fig.1c, compared to the unfiltered signal, there is an enhancement of 4 times on average.

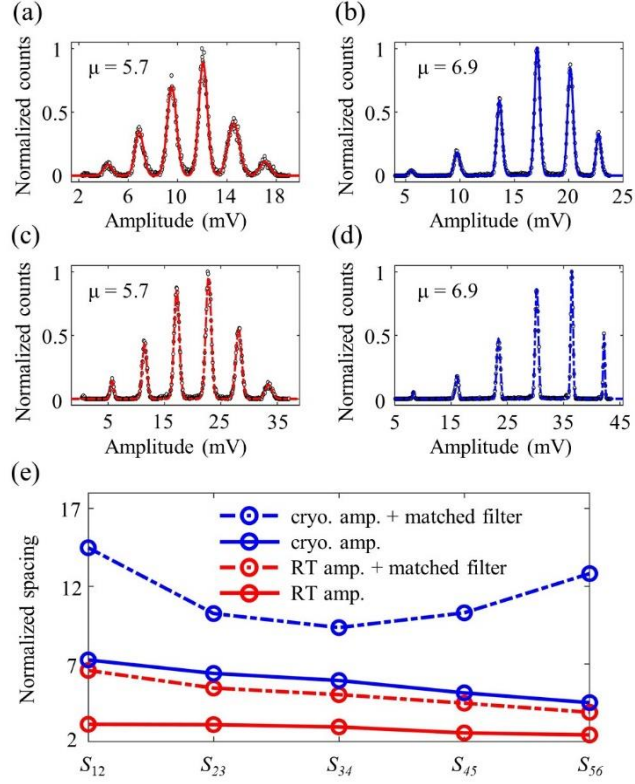

FIG. 2. Histograms of the pulse amplitude for different readout setups and signal processing methods. (a) and (b) are for an RT amplifier with and without a matched filter, respectively. The mean photon number is µ = 5.7. (c) and (d) are for a cryogenic amplifier with and without a matched filter, respectively. The mean photon number is µ = 6.9. The points are experimental data while the lines are fitting curves using multiple Gaussian functions. (e). The normalized amplitude spacing extracted from (a)~(d).

Benefiting from the enhancement of the SNR, the pulse amplitude distribution can be measured more accurately. After extracting the amplitude of peaks in the distribution, we can have the photon number distribution. To calculate the input mean photon number, probability of multiple photons hitting on a same pixel was taken into account. To better illustrating the performance of the matched filter, we used a cryogenic amplifier (CITLF1 from CMT) placed at 1.5 K to read out the 6-pixel PNR-SNSPD for comparison. Figure 2a and b show the pulse amplitude distributions using an RT amplifier and a cryogenic preamplifier, respectively. Figure 2c and d show the corresponding distributions after the matched filter. To quantify the improvement, we defined a normalized spacing between two adjacent peaks, which was calculated by

$$S_{n,n+1} = \frac{(p_{n+1}-p_n)}{2.355 \cdot (\sigma_{n+1}+\sigma_n)/2} \quad (5)$$

In Equ.5, the numerator gives the amplitude spacing between adjacent peaks, while the denominator is the average full-wave-at-half maximum (FWHM) of the two peaks assuming they follow a Gaussian distribution.

From Fig.2e, we can compare $S_{n,n+1}$ for different cases. When a RT amplifier (RF bay, LNA 650) was used, the normalized spacing was $S_{12}$ = 3.12, $S_{23}$ = 3.10, $S_{34}$ = 2.95, $S_{45}$ = 2.57 and $S_{56}$ = 2.44. By applying the match filter, the normalized spacing increased to $S_{12}$ = 6.59, $S_{23}$ = 5.45, $S_{34}$ = 5.03, $S_{45}$ = 4.48 and $S_{56}$ = 3.89. These values approached to the normalized spacing when a cryogenic amplifier was used, which were $S_{12}$ = 7.26, $S_{23}$ = 6.40, $S_{34}$ = 5.94, $S_{45}$ = 5.14 and $S_{56}$ = 4.51. If we applied the matched filter on the output from cryogenic readout, such values further increased to $S_{12}$ = 14.47, $S_{23}$ = 10.24, $S_{34}$ = 9.34, $S_{45}$ = 10.29 and $S_{56}$ = 12.81. As the output signal is proportional to $1/(N + \frac{Z_0}{R_s})$, it implies that by applying a matched filter and setting the minimum spacing to the FWHM (namely the minimum $S_{n,n+1}$ to be 1), we can estimate a maximum array size of 26 with an RT amplifier and a maximum array size of 65 with a cryogenic preamplifier.

From Fig.2e, we can see that the normalized spacing is not constant. One reason is that the output signal model contains nonlinear dependence on $n$ when a 50 Ω amplifier is used[19]. The increment of the signal becomes less as $n$ goes large Meanwhile, time constants in the bi-exponential curve for different $n$-photon detections have slight difference. Thus, a fixed wavelet in the matched filter cannot be optimal for all $n$-photon detections. Experimentally, we can first capture the average waveform and use this to design the optimal match filter for each $n$-photon detection to further increase the SNR.

Conventionally, pulse arrival time is measured by setting a voltage threshold on the rising edge. After the matched filter, since the maximum SNR happens at the peak, we used the time of the peak $t_{pn}$ for characterizing the arrival time of the detection pulse. By collecting a number of $t_{pn}$, we can have the distribution of the pulse arrival time, from which we can derive the timing jitter from the FWHM of the histogram. As shown in Fig.3, for all cases the timing jitter after the matched filter shows slight improvements. For single-photon detection, as shown in Fig.3a and b, the timing jitter reduced from 305 ps to 290 ps for a RT amplifier and 94 ps to 86 ps for a cryogenic amplifier. For multiphoton detection, since the SNR is better, the timing jitter affected by noise became less prominent. For five-photon detection, as shown in Fig.3c and d, the timing jitter reduced from 114 ps to 94 ps for a RT amplifier and 73 ps to 55 ps for a cryogenic amplifier.

It seems that the matched filter does not give much improvement on the timing jitter compared to the improvement on SNR. Although the $SNR_n$ at the peak is the maximum, in the presence of noise, determining the location of the peak is still difficult as the noise will contribute to a time variation by $\frac{\sigma}{\frac{dy_n}{dt}}$. At the peak value, $\frac{dy_n}{dt}\big|_{t=t_{pn}}$ approaches to zero, indicating that a small noise will convert to a large time variation in $t_{pn}$. For unfiltered pulses, the rising edge has a sharp transition, giving a large $\frac{dr_n}{dt}$. Thus, although the SNR is relatively worse, by setting a proper threshold on the rising edge where the slope is the maximum, the relative low timing jitter can still be obtained.

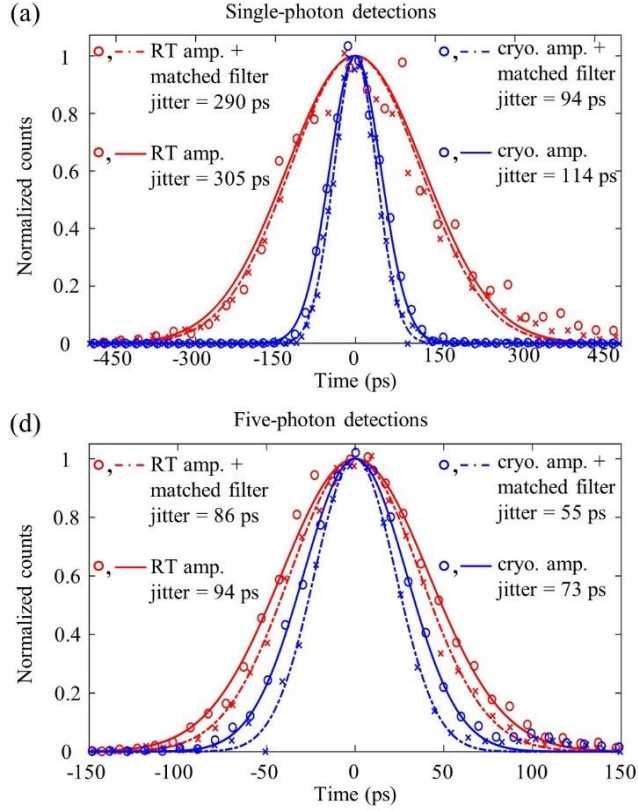

FIG. 3. Histograms of the normalized counts over delay time between the detected pulse and the reference of a femtosecond laser at different readout setups and signal processing methods. (a) The single-photon detections. (b) Five-photon detections. The points are experimental data and the lines are fitting curves with a Gaussian function.

The above analysis also suggests that the matched filter performed better than a low-pass filter. From the spectrum of pulses shown in Fig.1b, we already know that the spectrum SNR increases as frequency goes low. Thus, but setting a proper cut-off frequency, the SNR of a pulse in the time domain can increase. As shown in Fig.4a, the normalized separation $S_{12}$ reached a maximum of 13.2 when a cryogenic amplifier was used. At this maximum the cut-off frequency was 65 MHz, agreeing with the characteristic frequency $f_c$ from the pulse spectrum shown in Fig.1b. For pulses readout by an RT amplifier, the maximum of the normalized separation was 5.5 at a cut-off frequency of 30 MHz. The difference in the cut-off frequency was mainly due to the different noise figures and bandwidth of the amplifiers.

It is noticeable that for all cut-off frequencies, the normalized spacing values are lower than the values obtained after the matched filter, suggesting that the matched filter is the optimal filter in terms of SNR. The problem for low-pass filtering is that the timing jitter is seriously affected if the cut-off frequency is set too low This is clearly shown in Fig. 4b. For example, in the case of using a cryogenic amplifier, at the cut-off frequency of 65 MHz which gives the maximum spacing, the timing jitter is as large as 81 ps, while the time jitter is 55 ps with the matched filter.

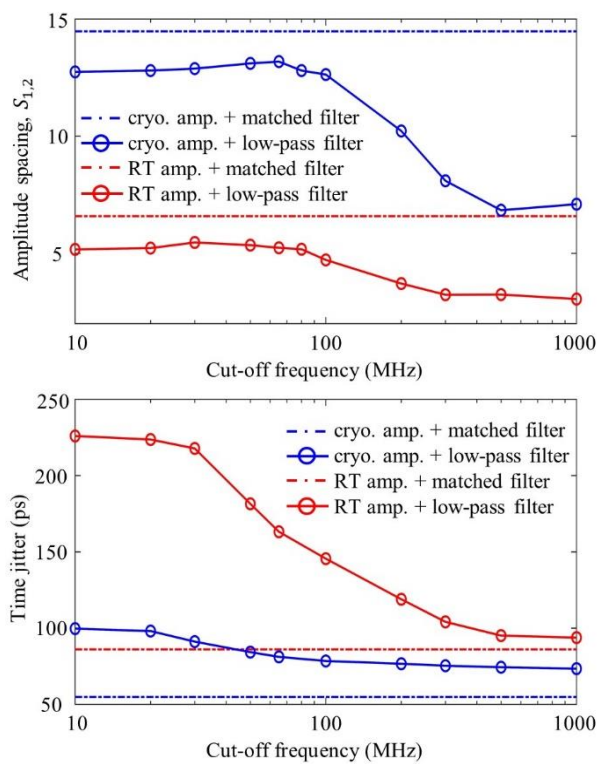

FIG. 4. The pulse amplitude spacing $S_{12}$ (a) and the timing jitter (b) for different readout setups with a low passing filter at the different cut-off frequencies. For better comparison, the values obtained after a matched filter are marked as the dash lines.

In conclusion, we have designed a digital matched filter for series PNR-SNSPDs and achieved a significant improvement of the SNR. With this signal processing, we observed a better separation of the pulse amplitude distribution. This also indicates that increasing the array size of the series PNR-SNSPD, namely the maximum resolved photon number, to 65 or even larger is possible. Compared to a low-pass filter, the matched filter can obtain the low timing jitter and high SNR simultaneously. In our current system, the signal processing was done after capturing waveforms from a fast oscilloscope. To enable real-time processing, the matched filter needs to be realized on a platform of high-speed FPGA (field-programmable gate array) and ADC.

**ACKNOWLEDGMENTS**

This work was supported by the National Natural Science Foundation (Nos. 62071214, 61801206, 61521001, 61571217, 61801209, and 11227904), the National Key R&D Program of China Grant (2017YFA0304002), the Program for Innovative Talents and Entrepreneur in Jiangsu, the Fundamental Research Funds for the Central Universities, the Priority Academic Program Development of Jiangsu Higher Education Institutions (PAPD), and the Jiangsu Provincial Key Laboratory of Advanced Manipulating Technique of Electromagnetic Waves.

# References

[1] R.H. Hadfield, Nat. Photonics **3**, 696 (2009).

[2] C.M. Natarajan, M.G. Tanner, and R.H. Hadfield, Supercond. Sci. Technol. **25**, (2012).

[3] P. Kok, W.J. Munro, K. Nemoto, T.C. Ralph, J.P. Dowling, and G.J. Milburn, Rev. Mod. Phys. **79**, 135 (2007).

[4] M. Cattaneo, M.G.A. Paris, and S. Olivares, Phys. Rev. A **98**, 1 (2018).

[5] F.E. Becerra, J. Fan, and A. Migdall, Nat. Photonics **9**, 48 (2014).

[6] X. Li, J. Tan, K. Zheng, L. Zhang, L. Zhang, W. He, P. Huang, H. Li, B. Zhang, Q. Chen, R. Ge, S. Guo, T. Huang, X. Jia, Q. Zhao, X. Tu, L. Kang, J. Chen, and P. Wu, Photonics Res. **8**, 637 (2020).

[7] L. Xue, Z. Li, L. Zhang, D. Zhai, Y. Li, S. Zhang, M. Li, L. Kang, J. Chen, P. Wu, and Y. Xiong, Opt. Lett. **41**, 3848 (2016).

[8] Y. Altmann, S. McLaughlin, M.J. Padgett, V.K. Goyal, A.O. Hero, and D. Faccio, Science (80-.). **361**, (2018).

[9] A. Divochiy, F. Marsili, D. Bitauld, A. Gaggero, R. Leoni, F. Mattioli, A. Korneev, V. Seleznev, N. Kaurova, O. Minaeva, G. Gol'Tsman, K.G. Lagoudakis, M. Benkhaoul, F. Lévy, and A. Fiore, Nat. Photonics **2**, 302 (2008).

[10] F. Marsili, D. Bitauld, A. Gaggero, S. Jahanmirinejad, R. Leoni, F. Mattioli, and A. Fiore, New J. Phys. **11**, (2009).

[11] S. Jahanmirinejad, G. Frucci, F. Mattioli, D. Sahin, A. Gaggero, R. Leoni, and A. Fiore, Appl. Phys. Lett. **101**, 072602 (2012).

[12] X. Tao, S. Chen, Y. Chen, L. Wang, X. Li, X. Tu, X. Jia, Q. Zhao, L. Zhang, L. Kang, and P. Wu, Supercond. Sci. Technol. **32**, (2019).

[13] F. Mattioli, Z. Zhou, A. Gaggero, R. Gaudio, R. Leoni, and A. Fiore, Opt. Express **24**, 9067 (2016).

[14] C. Wan, Z. Jiang, L. Kang, and P. Wu, IEEE Trans. Appl. Supercond. **27**, 1 (2017).

[15] M. Ejrnaes, R. Cristiano, O. Quaranta, S. Pagano, A. Gaggero, F. Mattioli, R. Leoni, B. Voronov, and G. Gol'Tsman, Appl. Phys. Lett. **91**, (2007).

[16] A.N. McCaughan and K.K. Berggren, Nano Lett. **14**, 5748 (2014).

[17] W. Guo, R.P. Gardner, and C.W. Mayo, Nucl. Instruments Methods Phys. Res. Sect. A Accel. Spectrometers, Detect. Assoc. Equip. **544**, 668 (2005).

[18] Q. Zhao, L. Zhang, T. Jia, L. Kang, W. Xu, J. Chen, and P. Wu, Appl. Phys. B Lasers Opt. **104**, 673 (2011).

[19] G.L. Turin, Ire Trans. Inf. Theory (1960).

[20] S. Jahanmirinejad and A. Fiore, Opt. Express **20**, 5017 (2012).